\documentclass[12pt, prd, showpacs]{revtex4}
\usepackage{amssymb}
\usepackage{amsmath}

\input{tcilatex}

\begin{document}

\title{On regular frames near rotating black holes}
\author{O. B. Zaslavskii}
\affiliation{Department of Physics and Technology, Kharkov V.N. Karazin National
University, 4 Svoboda Square, Kharkov 61022, Ukraine}
\affiliation{Institute of Mathematics and Mechanics, Kazan Federal University, 18
Kremlyovskaya St., Kazan 420008, Russia}
\email{zaslav@ukr.net }

\begin{abstract}
We consider the metric of a generic axially symmetric rotating stationary
black hole. The general approach is developed that enables us to construct
coordinate frame regular near the horizon. As explicit examples, the Kerr
and Kerr-Newmann-(anti-)de Sitter metrics are considered. It is shown how
the rotational versions of the Painlev\'{e}-Gullstrand and Doran coordinates
appear in this scheme as particular cases. For the 2+1 version of the metric
the direct generalization of the Lema\^{\i}tre coordinate system is
obtained. It is shown that the possibility of introducing a regular frame is
indirectly related to the constancy of a black hole angular velocity and the
rate with which the metric coefficient responsible for the rotation of
spacetime, tends to it.
\end{abstract}

\maketitle

\section{Introduction}

The metric of the Kerr black hole is well known \cite{kerr}. Nonetheless,
its description continues to attract attention \ since different coordinate
frames are relevant in different physical contexts. First of all, it
concerns necessity to have well-defined coordinates near the horizon where
the standard Boyer-Lindquist ones \cite{bl} fail. To this end, new
coordinate systems were suggested that can be considered as generalization
of the Painlev\'{e}-Gusstland coordinates from the Schwarzschild case or
their modification \cite{dor}, \cite{nat}. Meanwhile, in real astrophysical
circumstances black holes are surrounded by matter, so their metric may
differ from the pure Kerr solutions thus representing so-called "dirty"
black holes. This leads to the question of finding regular frames near quite
generic rotating black holes. Also, this problem is of obvious general
interest on its own right. In doing so, we face with two different but
tightly related problems: (i) how to describe coordinate transformations
that brings the metric into a regular form thus generalizing the procedure
found previously for the Kerr solution, (ii) for \ given metric, how to
unify different transformations, so that coordinate frames of Refs. \cite%
{dor}, \cite{nat} or their counterpart for dirty black holes be particular
cases of some general scheme.

Below, we describe approach that resolves both issues. We check it for the
Kerr and Kerr-Newman-(anti-)de Sitter metric and show that it agrees with
previous findings. We also demonstrate our approach for the rotational
analogue of the Bertotti-Robinson metric.

\section{Form of metric}

Let us consider the generic stationary axially symmetric metric. Its form
can be written as%
\begin{equation}
ds^{2}=-N^{2}dt^{2}+g_{\phi }(d\phi -\omega dt)^{2}+dn^{2}+g_{zz}dz^{2}\text{%
,}  \label{m1}
\end{equation}%
where all metric coefficient do not depend on $t$ and $z$ (see, e.g. Sec. II
and especially eq. (11) in Ref. \cite{vis}). Introducing a new variable
according to $dn=\frac{dr}{A(r)}$, one can rewrite it in the form%
\begin{equation}
ds^{2}=-N^{2}dt^{2}+g_{\phi }(d\phi -\omega dt)^{2}+\frac{dr^{2}}{A(r)}%
+g_{zz}dz^{2}\text{.}  \label{met2}
\end{equation}%
Here, $A=A(r)$ does not depend on $z$. In principle, it is sufficient to
consider further the metric in the form (\ref{met2}). However, the famous
Kerr metric and its extension to the presence of the electric charge and
cosmological constant do not have such a form in the Boyer-Lindquiste
coordinates or their generalization. To handle properly relevant
generalizations of \ the Kerr metric, we will consider somewhat different
form

\begin{equation}
ds^{2}=-N^{2}dt^{2}+g_{\phi }(d\phi -\omega dt)^{2}+\frac{dr^{2}}{A(r,\theta
)}+g_{\theta }d\theta ^{2},  \label{m}
\end{equation}%
in which the dependence on $\theta $ is allowed in . Also, we use the angle
variable $\theta $ instead of $z$.

We assume that%
\begin{equation}
N^{2}=\alpha \Delta _{r}\text{, }A=\frac{\Delta _{r}}{\rho ^{2}}\text{,}
\label{abn}
\end{equation}%
where $\Delta _{r}$ does not depend on $\theta $, $\alpha $ and $\rho $
depend on both $r$ and $\theta $. The maximum zero $r=r_{+}$ of $\Delta _{r}$
corresponds to the event horizon. It is supposed that $\alpha $ and $\rho $
remain finite and nonzero at $r=r_{+}$. Then, the metric can be written as

\begin{equation}
ds^{2}=-N^{2}dt^{2}+g_{\phi }(d\phi -\omega dt)^{2}+\frac{\rho ^{2}}{\Delta
_{r}}dr^{2}+g_{\theta }d\theta ^{2}.  \label{aro}
\end{equation}

This form can be substantiated on the basis of regularity. Indeed, according
to the zero law of black hole mechanics \cite{wald}, the surface gravity

\begin{equation}
\kappa =\lim_{r\rightarrow r_{+}}\sqrt{(\nabla N)^{2}}
\end{equation}%
should be finite and constant on the event horizon.

For nonextremal black holes, by definition,%
\begin{equation}
N^{2}\sim r-r_{+}\text{.}
\end{equation}%
Let us admit for a moment that the horizon corresponds to $r_{+}(\theta )$.
Then, the term $g^{\theta \theta }(\frac{\partial N}{\partial \theta })^{2}$
would lead to divergency of the quantity under consideration. To have
regular horizon, we exclude the dependence of $\Delta _{r}$ on $\theta $.
For extremal horizons, these arguments do not work. However, more refined
examination shows that the dependence on $\theta $ should be excluded in
this case as well (see Sec. III C. 2 in \cite{dirty}). It is worth noting
that the form (\ref{aro}) with $\Delta _{r}=\Delta _{r}(r)$ is typical of
the Kerr, Kerr-Newman-(anti)de Sitter metrics.

\section{Coordinate transformations}

Near the horizon the coordinate system (\ref{aro}) fails. Our goal is to
find a new one regular in the vicinity of $r_{+}$. To this end, let us make
the transformation to new (barred) coordinates according to

\begin{equation}
dt=d\bar{t}+\frac{zdr}{\Delta _{r}}\text{,}  \label{t}
\end{equation}%
\begin{equation}
d\phi =d\bar{\phi}+\frac{\xi dr}{\Delta _{r}}+\delta d\theta ,\text{ }r=\bar{%
r}\text{.}  \label{phi}
\end{equation}%
In principle, one can add the term $\gamma d\theta $ in (\ref{t}) but, to
avoid unnecessary complication, we put $\gamma =0$. This holds, in
particular, in the Painlev\'{e}-Gullstrand and Doran coordinate systems in
the Kerr metric \cite{nat}.

As $dt$ should be a total differential, we require that $z$ does not depend
on $\theta $, $z=z(r)$.

Now we have%
\begin{equation}
d\phi -\omega dt=d\bar{\phi}-\omega d\bar{t}+hdr+d\theta \delta ,
\end{equation}%
\begin{equation}
g_{rr}=\mu +g_{\phi }h^{2}\text{, }
\end{equation}%
where by definition%
\begin{equation}
h\Delta _{r}=\xi -\omega z\text{,}  \label{2}
\end{equation}%
\begin{equation}
\mu \Delta _{r}=\rho ^{2}-z^{2}\alpha \text{,}  \label{ksi}
\end{equation}%
where we tok into account (\ref{abn}). Then,%
\begin{equation}
ds^{2}=-N^{2}d\bar{t}^{2}-2z\alpha drd\bar{t}+g_{\phi }(d\bar{\phi}-\omega d%
\bar{t}+hdr+d\theta \delta )^{2}+\mu dr^{2}+d\theta ^{2}g_{\theta }
\label{m2}
\end{equation}%
We choose the functions $z$, $h$ to kill divergences in the metric
coefficient $g_{rr}$. To this end, we require $h$ and $\mu $ to be bounded
on the horizon.

We have from (\ref{2}) and (\ref{ksi}) that%
\begin{equation}
\xi =\omega z+h\Delta _{r}\text{,}  \label{ks}
\end{equation}%
\begin{equation}
\mu =\frac{\rho ^{2}-z^{2}(r)\alpha }{\Delta _{r}}.  \label{mf}
\end{equation}%
The finiteness of $h$ and $\mu $ on the horizon entails the necessity of the
conditions%
\begin{equation}
(\xi -\omega z)_{r=r_{+}}=0\text{,}  \label{omhor}
\end{equation}%
\begin{equation}
\left( \rho ^{2}-z^{2}\alpha \right) _{r=r_{+}}=0\text{.}  \label{za}
\end{equation}%
This implies that $\frac{\rho ^{2}}{\alpha }$ does \ not depend on $\theta $
on the horizon. Also, the integrability condition in (\ref{phi}) requires%
\begin{equation}
\delta _{,r}=\left( \frac{\xi }{\Delta _{r}}\right) _{,\theta }\text{.}
\label{int2}
\end{equation}

We can rewrite (\ref{m2}) in the form%
\begin{equation}
ds^{2}=-\frac{\alpha \rho ^{2}}{\mu }d\bar{t}^{2}+\mu (dr-Vd\bar{t}%
)^{2}+g_{\phi }(d\bar{\phi}-\omega d\bar{t}+hdr+d\theta \delta )^{2}+d\theta
^{2}g_{\theta }\text{,}  \label{mu}
\end{equation}%
where%
\begin{equation}
V=\frac{\alpha z}{\mu }\text{.}  \label{V}
\end{equation}

Thus we see that specifying the functions $z(r),$and $h$ we obtain a class
of coordinate transformation that make the metric regular near the horizon.
Other two functions $\xi $ and $\mu $ can be found from (\ref{ks}), (\ref{mf}%
). Below, we specify some natural choices and analyze important particular
cases.

\section{Regular frames and rigid rotation of the horizon}

There are two typical cases. If $h=0$, it follows from (\ref{2}) that%
\begin{equation}
\xi =\omega z\text{.}  \label{yz}
\end{equation}%
This is the generalization of rotational version of the Painlev\'{e}%
-Gillstrand coordinate frame introduced in \cite{nat} for the Kerr metric.
Now, eq. (\ref{int2}) reduces to%
\begin{equation}
\delta _{,r}=\frac{z}{\Delta _{r}}\omega _{,\theta }\text{.}  \label{int3}
\end{equation}

If the horizon is regular in the sense that tetrad component of the
curvature tensor measured by free-falling observers remain bounded, the
quantity%
\begin{equation}
\omega _{,\theta }=O(N^{2})  \label{om2}
\end{equation}%
near the horizon. This was shown in \cite{dirty}, see Tables I and II there.
Then, it is clear from (\ref{int3}) that $\delta _{,r}$ is finite near the
horizon and $\delta $ can be found by direct integration. As $\delta $ is
finite, frame (\ref{mu}) is regular near the horizon. This agrees with the
previous consideration done for the Kerr metric in \cite{nat}.

The case $\delta =0$ is generalization of the Doran system \cite{dor}. In
doing so, the function $\xi $ should depend on $r$ only since $d\phi $ (\ref%
{phi}) has to be exact differential. Then, it follows from (\ref{omhor})
that on the horizon, where $\omega $ has the meaning of a black hole angular
velocity $\omega _{H}$, this angular velocity must not depend on $\theta $.
But this holds true for regular black holes automatically \cite{bch} and
agrees with (\ref{om2}).

Thus we see that the regularity of a new coordinate frame not only requires
the finiteness of corresponding functions $\delta $ and $h$ but, indirectly,
relies on rather subtle properties inherent to regular horizons, whatever
frame be used. More precisely, the possibility of introducing the frames
under discussion is related to the constancy of a black hole angular
velocity $\omega _{H}$ and, moreover, may depend on the rate with which $%
\omega $ tends to $\omega _{H}$.

\section{Particular class of metrics}

The formulas are simplified if one can take%
\begin{equation}
\mu =\alpha \rho ^{2}\text{.}  \label{mc}
\end{equation}%
Then, it follows from (\ref{ksi}), (\ref{V}) that%
\begin{equation}
V=\frac{\sqrt{1-N^{2}}}{\rho \sqrt{\alpha }}=\frac{z}{\rho ^{2}}\text{,}
\label{v}
\end{equation}%
where the combination%
\begin{equation}
\frac{\sqrt{1-N^{2}}\rho }{\sqrt{\alpha }}=z(r)  \label{com}
\end{equation}%
should depend on $r$ only.

\subsection{Choice 1}

\begin{equation}
h=0\text{.}
\end{equation}%
Taking into account (\ref{mc}), we obtain 

\begin{equation}
ds^{2}=-d\bar{t}^{2}+\alpha \rho ^{2}(dr-\frac{z}{\rho ^{2}}d\bar{t}%
)^{2}+g_{\phi }(d\bar{\phi}-\omega d\bar{t}+\delta d\theta )^{2}+g_{\theta
}d\theta ^{2}\text{.}  \label{ch1}
\end{equation}%
This can be called the generalized Painlev\'{e}-Gullstrand coordinate frame.

According to (\ref{int2}) and (\ref{yz}),%
\begin{equation}
\delta _{,r}=\left( \frac{\omega z}{\Delta _{r}}\right) _{,\theta }=\frac{%
z\omega _{,\theta }}{\Delta _{r}}\text{.}  \label{i1}
\end{equation}%
since $z$ and $\Delta _{r}$ are functions of $r$ only.

\subsection{Choice 2}

Let us choose

\begin{equation}
\delta =0\text{.}
\end{equation}%
In particular, we can take also%
\begin{equation}
h=\frac{\omega \rho ^{2}}{z},
\end{equation}%
then%
\begin{equation}
\xi =\frac{\omega \rho ^{2}}{\alpha z}=\frac{h}{\alpha }\text{.}  \label{n2}
\end{equation}%
Then, with (\ref{2}) taken into account we have from (\ref{mu})%
\begin{equation}
ds^{2}=-d\bar{t}^{2}+g_{rr}(dr-\frac{z}{\rho ^{2}}d\bar{t}+\frac{g_{\phi }}{%
g_{rr}}\frac{\omega \rho ^{2}}{z}d\bar{\phi})^{2}+g_{\phi }(1-\frac{\omega
^{2}\rho ^{4}}{z^{2}}\frac{g_{\phi }}{g_{rr}})d\bar{\phi}^{2}+g_{\theta
}d\theta ^{2}\text{,}  \label{dorg}
\end{equation}%
where%
\begin{equation}
g_{rr}=\alpha \rho ^{2}+\frac{\rho ^{4}\omega ^{2}}{z^{2}}g_{\phi }\text{.}
\label{rr}
\end{equation}

This is generalization of the Doran coordinate system \cite{dor}.

One important reservation is in order. The above choice (\ref{mc}) implies
that the left hand side of eq. (\ref{com}) is a function of $r$ only. This
is the case for the Kerr metric (see below) but, in general, may or may not
be valid. In particular, we will see that for the Kerr-Newman-(anti-)de
Sitter metric, another, more complicated ansatz works successfully instead
of (\ref{mc}).

\section{Kerr metric}

For the Kerr metric in the Boyer-Lindquiste coordinates,%
\begin{equation}
ds^{2}=-\frac{\Delta _{r}}{\rho ^{2}}(dt-a^{2}\sin ^{2}\theta d\phi )^{2}+%
\frac{\sin ^{2}\theta }{\rho ^{2}}[adt-(r^{2}+a^{2})d\phi ]^{2}+\frac{\rho
^{2}dr^{2}}{\Delta _{r}}+\rho ^{2}d\theta ^{2},  \label{kerr}
\end{equation}%
where $M$ is a black hole mass, 
\begin{equation}
\rho ^{2}=r^{2}+a^{2}\cos ^{2}\theta ,  \label{ro}
\end{equation}%
$a$ is the angular momentum per unit mass.

Then, the metric (\ref{kerr}) is rewritten in the form (\ref{m}),

\begin{align}
\omega & =\frac{2Mra}{\Sigma \rho ^{2}},  \label{wnk} \\
N^{2}& =\frac{\Delta }{\Sigma }\text{,}  \label{nk}
\end{align}%
\begin{equation}
\alpha =\frac{1}{\Sigma }\text{,}  \label{al}
\end{equation}%
\begin{equation}
\Sigma =r^{2}+a^{2}+\frac{2Mra^{2}}{\rho ^{2}}\sin ^{2}\theta \text{,}
\end{equation}%
\begin{equation}
\Delta =r^{2}-2Mr+a^{2}.  \label{dk}
\end{equation}%
By substitution into (\ref{v}) one finds%
\begin{equation}
z=\sqrt{2Mr(r^{2}+a^{2})}\text{.}
\end{equation}%
One also finds that for choice 2 the expression (\ref{n2}) reads 
\begin{equation}
\xi =\frac{2Mra}{\sqrt{2Mr(r^{2}+a^{2})}}\text{,}
\end{equation}%
so $\xi $ is a \ function of $r$ only and condition (\ref{int2}) is
satisfied.

\section{Kerr-Newman-(anti-)de Sitter}

The metric has the form (see, e.g. Sec. 11.3 in \cite{exact})%
\begin{equation}
ds^{2}=-\frac{\Delta _{r}}{\Xi ^{2}\rho ^{2}}(dt-a\sin ^{2}\theta d\phi
)^{2}+\frac{\Delta _{\theta }\sin ^{2}\theta }{\Xi ^{2}\rho ^{2}}%
[adt-(r^{2}+a^{2})d\phi ]^{2}+\frac{\rho ^{2}dr^{2}}{\Delta _{r}}+\frac{\rho
^{2}d\theta ^{2}}{\Delta _{\theta }},
\end{equation}%
where%
\begin{equation}
\Xi =1+\frac{\Lambda a^{2}}{3},
\end{equation}%
\begin{equation}
\Delta _{r}=(r^{2}+a^{2})(1-\frac{\Lambda r^{2}}{3})-2Mr+e^{2}\text{,}
\end{equation}%
\begin{equation}
\Delta _{\theta }=1+\frac{\Lambda a^{2}}{3}\cos ^{2}\theta \text{, }
\end{equation}%
$\Lambda $ is the cosmological constant, $e$ being the electric charge, $%
\rho $ is given by (\ref{ro}). Then, one finds that%
\begin{equation}
N^{2}=\frac{\Delta _{r}\Delta _{\theta }}{\Xi ^{2}\Sigma }\text{,}
\end{equation}%
\begin{equation}
\alpha =\frac{\Delta _{\theta }}{\Xi ^{2}\Sigma }\text{,}
\end{equation}%
\begin{equation}
\omega =a\frac{(r^{2}+a^{2})\Delta _{\theta }-\Delta _{r}}{\rho ^{2}\Sigma }%
\text{,}  \label{oms}
\end{equation}%
where%
\begin{equation}
\Sigma =\frac{(r^{2}+a^{2})^{2}}{\rho ^{2}}\Delta _{\theta }-\frac{a^{2}\sin
^{2}\theta \Delta _{r}}{\rho ^{2}}\text{.}  \label{rs}
\end{equation}

It follows from (\ref{mf}) that now 
\begin{equation}
\mu =\frac{\rho ^{2}\Sigma \Xi ^{2}-z^{2}\Delta _{\theta }}{\Delta
_{r}\Sigma \Xi ^{2}}\text{.}
\end{equation}

It is natural to use the ansatz%
\begin{equation}
z=\Xi (r^{2}+a^{2})k(r)\text{,}  \label{zs}
\end{equation}%
where $k(r)$ is a new function. Then,%
\begin{equation}
\mu =\frac{(r^{2}+a^{2})^{2}\Delta _{\theta }-a^{2}\sin ^{2}\theta \Delta
_{r}-\Delta _{\theta }(r^{2}+a^{2})^{2}k^{2}}{\Delta _{r}\Sigma }\text{.}
\end{equation}%
Let us write%
\begin{equation}
k^{2}=1-\frac{\Delta _{r}}{f^{2}}\text{,}
\end{equation}%
where $f\neq 0$ is finite for $\Delta _{r}=0$, so that $k=1$ on the horizon.
This ensures the finiteness of $\mu $ at $\Delta _{r}=0$. Then, (\ref{t})
takes the form%
\begin{equation}
dt=d\bar{t}+\frac{\Xi (r^{2}+a^{2})k(r)}{\Delta _{r}}dr\text{.}
\end{equation}%
This corresponds to the first part of eq. (5) in \cite{lin} (with another
choice of signs).

In \cite{lin}, the choice%
\begin{equation}
\xi =a\Xi k  \label{ns}
\end{equation}%
was made. Then, we see that according to (\ref{2}), (\ref{oms}), (\ref{zs}) 
\begin{equation}
h=\frac{\Xi ak}{\Sigma }\text{,}  \label{hs}
\end{equation}%
so $h$ is indeed finite on the horizon. Meanwhile, instead of (\ref{ns}), (%
\ref{hs}) one can take a more general expression according to (\ref{2})
where $h$ is a function of $r$ and $\theta $ finite on the horizon,
otherwise arbitrary.

\section{Rotational analogue of Bertotti-Robinon spacetime}

In this section, we consider one more example of rotating metrics with the
horizon. This is the analogue of the Bertotti-Robinson metric. For the first
time, such a class of metrics was found in \cite{hm}. It was shown that such
metrics appear naturally as the result of the limiting transition from the
nonextremal metric to its extremal state when a black hole is enclosed in a
cavity. In doing so, in the canonical thermal ensemble the temperature of a
system is kept fixed on a boundary. Another version of such metrics
corresponding to the extremal horizon was found in \cite{bh}. For the sake
of definiteness, we choose the variant corresponding to eq. (2.6) of \cite%
{bh}. The metric takes the form%
\begin{equation}
ds^{2}=\frac{(1+\cos ^{2}\theta )}{2}[-\frac{r^{2}}{r_{0}^{2}}dt^{2}+\frac{%
r_{0}^{2}}{r^{2}}dr^{2}+r_{0}^{2}d\theta ^{2}]+\frac{2r_{0}^{2}\sin
^{2}\theta }{1+\cos ^{2}\theta }(d\phi +\frac{r}{r_{0}^{2}}dt)^{2}\text{,}
\label{th}
\end{equation}%
where $r_{0}^{2}=2M^{2}$, $M$ being the mass of the original Kerr black hole
from which by means of the limiting transition the metric (\ref{th}) is
obtained. In our terms, 
\begin{equation}
\Delta _{r}=r^{2}\text{, }\alpha =\frac{(1+\cos ^{2}\theta )}{2r_{0}^{2}}%
\text{, }\rho ^{2}=\frac{(1+\cos ^{2}\theta )r_{0}^{2}}{2}\text{, }\omega =-%
\frac{r}{r_{0}^{2}}\text{,}
\end{equation}

Near the horizon, the coordinate frame (\ref{th}) fails. Applying our
approach, we can choose%
\begin{equation}
z=r_{0}^{2}\sqrt{1-\frac{r^{2}}{r_{0}^{2}}}\text{,}
\end{equation}%
\begin{equation}
\mu =\frac{(1+\cos ^{2}\theta )}{2}\text{.}
\end{equation}%
Further, we can consider choices 1 and 2 from the above.

\subsection{\protect\bigskip Choice 1}

\begin{equation}
h=-\frac{r}{r_{0}^{2}}\frac{(1+\cos ^{2}\theta )}{2\sqrt{1-\frac{r^{2}}{%
r_{0}^{2}}}}\text{,}
\end{equation}%
\begin{equation}
\xi =-\frac{r}{r_{0}^{2}}z+hr^{2}\text{.}
\end{equation}%
As now $\omega _{,\theta }=0$, it is clear from (\ref{i1}) that we can also
take $\delta =0$.

According to (\ref{ch1}),

\begin{equation}
ds^{2}=-d\bar{t}^{2}+\frac{(1+\cos ^{2}\theta )^{2}}{4}(dr-\frac{2\sqrt{1-%
\frac{r^{2}}{r_{0}^{2}}}}{1+\cos ^{2}\theta }d\bar{t})^{2}+\frac{%
2r_{0}^{2}\sin ^{2}\theta }{1+\cos ^{2}\theta }(d\bar{\phi}+\frac{r}{%
r_{0}^{2}}d\bar{t})^{2}+\frac{(1+\cos ^{2}\theta )}{2}r_{0}^{2}d\theta ^{2}%
\text{.}
\end{equation}

\subsection{Choice 2}

\begin{equation}
\xi =-\frac{r}{r_{0}^{2}}z
\end{equation}%
\begin{equation}
h=0
\end{equation}

Eq. (\ref{dorg}) looks somewhat cumbersome, so we list only the metric
component (\ref{rr})%
\begin{equation}
g_{rr}=\frac{(1+\cos ^{2}\theta )^{2}}{4}[1+\frac{2r^{2}\sin ^{2}\theta }{%
(r_{0}^{2}-r^{2})(1+\cos ^{2}\theta )}].
\end{equation}

Thus, we successfully remove coordinate singularities.

\section{2+1 systems}

One more class of metrics for which the approach under discussion works are
the metrics in 2+1 gravity. This includes famous BTZ black hole \cite{btz}
and its generalization if matter surrounding the horizon is allowed. Now the
original metric in its general form can be obtain by discarding the last
term in (\ref{m}), so

\begin{equation}
ds^{2}=-N^{2}dt^{2}+g_{\phi }(d\phi -\omega dt)^{2}+\frac{dr^{2}}{A}.
\label{21}
\end{equation}

Now, the coefficients are supposed to depend on a variable $r$ only, so
integrability conditions imposed on functions $z$ and $\xi $ are satisfied
automatically.

As an example, we consider the BTZ black hole. Then, $A=N^{2}=\Delta _{r}$,%
\begin{equation}
\rho ^{2}=1=\alpha \text{, }g_{\phi }=r^{2}\text{,}
\end{equation}%
\begin{equation}
\omega =\frac{J}{2r^{2}}\text{,}
\end{equation}%
\begin{equation}
N^{2}=-M+\frac{r^{2}}{l^{2}}+\frac{J^{2}}{4r^{2}}\text{.}
\end{equation}%
Here, $M$ is the mass, $J$ being the angular momentum, the cosmological
constant $\Lambda =-\frac{1}{l^{2}}$. The choice $\mu =1$ is quite
sufficient. Then, it follows from (\ref{ksi}) that%
\begin{equation}
z=\sqrt{1-N^{2}}\text{.}
\end{equation}%
Now, the spacetime is not asymptotically flat, $N$ grows with $r$, so our
new coordinate frame is not applicable when $N^{2}>1$. However, this does
not cause any problem for the vicinity of the horizon that is just the
object of our interest. Applying the previous formulas, one finds for choice
1%
\begin{equation}
ds^{2}=-d\bar{t}^{2}+(dr-zd\bar{t})^{2}+r^{2}(d\bar{\phi}-\omega d\bar{t}%
)^{2}
\end{equation}%
and

\begin{equation}
ds^{2}=-d\bar{t}^{2}+g_{rr}(dr-zd\bar{t}+\frac{J^{2}}{2g_{rr}z}d\bar{\phi}%
)^{2}+\frac{r^{2}}{1+\frac{J^{2}}{4z^{2}r^{2}}}d\bar{\phi}^{2}\text{,}
\end{equation}%
\begin{equation}
g_{rr}=1+\frac{J^{2}}{4z^{2}r^{2}}
\end{equation}%
for choice 2.

The approach under discussion applies also to 3+1 metrics if the $\theta $
coordinate is suppressed, so effectively the metric reveals itself as 2+1
dimensional. For example, this happens for the description of particle
motion within the equatorial plane $\theta =\frac{\pi }{2}$.

\section{Rotational analogue of Lema\^{\i}tre frame}

Let us consider the 2+1 counterpart of (\ref{ch1}). Now, the factor $\rho
^{2}$ is a function of $r$ only, so it can be absorbed by $\Delta _{r}$ and
we may put $\rho =1$. Introducing a new variable according to $d\bar{r}=%
\sqrt{\alpha }dr$, one can cast it in the form%
\begin{equation}
ds^{2}=-d\bar{t}^{2}+(d\bar{r}-\bar{v}d\bar{t})^{2}+g_{\phi }(d\bar{\phi}%
-\omega d\bar{t})^{2}\text{,}
\end{equation}%
or 
\begin{equation}
ds^{2}=-dt^{2}f+\frac{d\bar{r}^{2}}{f}+g_{\phi }(d\bar{\phi}-\omega d\bar{t}%
)^{2}\text{,}
\end{equation}%
where 
\begin{equation}
\bar{v}=z\sqrt{\alpha },
\end{equation}
\begin{equation}
dt=d\bar{t}+\frac{\bar{v}}{f}d\bar{r},
\end{equation}%
\begin{equation}
f=1-\bar{v}^{2}\text{.}
\end{equation}

Making the transformation

\begin{equation}
\chi =t+\int \frac{d\bar{r}}{f\bar{v}}\text{,}
\end{equation}%
\begin{equation}
\tau =t+\int \frac{d\bar{r}\bar{v}}{f}\text{,}
\end{equation}%
one can obtain%
\begin{equation}
ds^{2}=-d\tau ^{2}+z^{2}\alpha d\chi ^{2}+g_{\phi }[d\bar{\phi}-\omega d\tau 
\frac{1}{f}+\frac{2\omega z^{2}\alpha d\chi }{f}]^{2}  \label{lem}
\end{equation}

It is easy to show from the equations of motions that coordinate lines $\chi
=const$, $\bar{\phi}=const$ are geodesics corresponding to particles with
the energy $E=m$ and angular momentum $L=0$ falling from infinity towards a
black hole. The relation between such a generalized Lema\^{\i}tre system and
the rotational version of the Painlev\'{e}-Gullstrand one is extension of
the similar relationship valid in the spherically symmetric case \cite{3}.

\section{Conclusions}

We considered a quite generic rotating axially symmetric stationary black
hole. We have described a general approach that enables one to remove
coordinate singularities that are present initially in the standard
Boyer-Lindquiste type of coordinates. Our approach involves two functions $%
z(r)$, $h(r,\theta ),$ that obey two equations (\ref{2}), (\ref{ksi}) plus
the conditions of finiteness on the horizon (\ref{za}). There is also one
more function $\delta $ that must obey the integrability condition (\ref%
{int2}). Thus there is a rather large freedom in the choice of a regular
coordinate frame. It is traced how some already known coordinate systems for
particular metrics (such as the Kerr or Kerr-Newman - (anti-)de Sitter one
and the vacuum throat metric) appear as particular cases. The system found
in \cite{dor} corresponds to $\delta =0$, the generalization of the Painlev%
\'{e}--Gullstrand system implies that $h=0$. As a result, a unified picture
that encompasses previously known frames is constructed. For the 2+1 systems
or problems in which 3+1 metric effectively reduces to 2+1 one, the
generalization of the Lema\^{\i}tre coordinate system is constructed. It is
related to the generalized version of the Painlev\'{e}-Gullstrand frame in a
quite simple way.

As coordinates frames under discussion are extendable across the horizon,
the formalism developed can be of some use for further investigation of
properties of near-horizon orbits \cite{ted11}, \cite{near}, including
nonequatorial motion. It can also help with studying the geometry inside a
black hole, for example in issues related to the measurement of a black hole
volume \cite{par}. One more issue is generalization of river-models \cite%
{riv} to the case of "dirty" black holes.

Our consideration reveals also an additional interesting property. It turned
out that between the possibility to introduce regular coordinate frames near
the horizon and the behavior of the coefficient $\omega $ near the horizon
(in particular, the constancy of the black hole angular velocity) there is
deep connection.

\begin{acknowledgments}
This work was funded by the subsidy allocated to Kazan Federal University
for the state assignment in the sphere of scientific activities. O. Z. also
thanks for support SFFR, Ukraine, Project No. 32367.
\end{acknowledgments}

\end{document}